\documentclass[aps,prl,twocolumn,groupedaddress,showpacs]{revtex4-1}

\usepackage{verbatim}
\usepackage{graphicx}
\usepackage{subfig}
\usepackage{dcolumn}
\usepackage{bm}
\usepackage{xcolor,soul}
\usepackage{amsthm} 
\usepackage{mathrsfs}
\usepackage{mathtools}

\newcommand{\braket} [2] {\langle{#1}|{#2}\rangle}

\begin{document}

\title{Self-energies in itinerant magnets: A focus on Fe and Ni}

\author{Lorenzo Sponza}
\affiliation{King's College London, London WC2R 2LS, UK}

\author{{Paolo Pisanti}}
\affiliation{King's College London, London WC2R 2LS, UK}

\author{Alena Vishina}
\affiliation{King's College London, London WC2R 2LS, UK}

\author{Swagata Acharya}
\affiliation{Indian Institute of Technology, Kharagpur 721302, India}
\affiliation{King's College London, London WC2R 2LS, UK}

\author{Dimitar Pashov}
\affiliation{King's College London, London WC2R 2LS, UK}

\author{Cedric Weber}
\affiliation{King's College London, London WC2R 2LS, UK}

\author{Julien Vidal}
\affiliation{EDF and R\&{}D, Department EFESE, 6 Quai Watier, 78401 Chatou, France}

\author{Gabriel Kotliar}
\affiliation{Rutgers University, New Brunswick, NJ, USA}

\author{Mark van Schilfgaarde}
\affiliation{King's College London, London WC2R 2LS, UK}

\date{\today}

\begin{abstract}

We present a detailed study of local and non-local correlations in the electronic structure of elemental transition metals carried out by means of the Quasiparticle Self-consistent GW (QS\emph{GW}) and Dynamical Mean Field Theory (DMFT).
Recent
high resolution ARPES and Haas-van Alphen data of two typical transition
metal systems (Fe and Ni) are used as 
case study.
(i) We find that the properties of Fe are very well described by QS\emph{GW}.
Agreement with cyclotron and very clean ARPES measurements is excellent, provided
that final-state scattering is taken into account.  This
establishes the exceptional reliability of QS\emph{GW} also in metallic systems.
(ii) Nonetheless QS\emph{GW} alone is not able
to provide an adequate description of the Ni ARPES data due to 
strong local spin fluctuations.
We surmount this deficiency by combining nonlocal charge fluctuations in
QS\emph{GW} with local spin fluctuations in DMFT (QS\emph{GW}+``Magnetic DMFT'').
(iii) Finally we show that the dynamics of the local fluctuations are actually not crucial.
The addition of an external static field can lead to
similarly good results if non-local correlations are included through QS\emph{GW}.

\end{abstract}


\maketitle

High-resolution spectroscopy is limited in transition metals, in
part because it is difficult to make sufficiently high
quality samples.  Fe and Ni are elements of which high quality films
have been grown, and
high-resolution angle-resolved photoemission spectroscopy (ARPES)
performed~\cite{Schafer05}. 
These experiments provides a good reference to test the validity of
different approximations of the electronic structure.

There are also not many calculations of spectral functions in
these materials.  Fe has been studied in the local-density
approximation (LDA)~\cite{callaway-wang_prb1977} and with corrections through
Dynamical-Mean Field Theory (DMFT)~\cite{LichtensteinFe09}. It is
not surprising that the LDA does not track the ARPES
experiment well \cite{Walter10}, but it has been found that
LDA+DMFT also fails to properly account for ARPES
data~\cite{LichtensteinFe09}.  The \emph{GW}
approximation~\cite{hedin_jpcm1999} is widely applied to many
kinds of insulators, but how well it describes 3$d$ transition
metals is much less established.

Through quasiparticle self-consistency (QS\emph{GW}) one determines the
noninteracting Green's function $G_{0}$ which is minimally distant
from the true Green's function $G$ \cite{Kotani07,Ismail14}.
Within QS\emph{GW} many electronic properties are in excellent agreement with experiment~\cite{Kotani07},
most notably the quasiparticle band structures.
Moreover, at self-consistency the poles of QS\emph{GW} $G_0(\mathbf{k},\omega)$ coincide with
the peaks in $G(\mathbf{k},\omega)$.
This means that there is
no many-body ``mass renormalization'' of the
noninteracting Hamiltonian, which allows for a direct association of
QS\emph{GW} energy
bands $E(\bf{k})$ with peaks in the spectral function
$A(\bf{k},\omega)$.
Thus, QS\emph{GW} provides an optimum framework to test the range
of validity, and the limitations to the \emph{GW} approximation.

In this work, we compare QS\emph{GW} results
to various experimental data in elemental
3$d$ materials in the Fermi liquid (FL) regime, with a heavy
focus on Fe because of the high quality of ARPES \cite{Schafer05}
and de Haas-van Alphen (dHvA)~\cite{Baraff73,Gold74} data available.
We will show that QS\emph{GW} and ARPES spectral functions
agree to within experimental resolution, with the proviso that
the final state scattering is properly accounted in interpreting the experimental data.
By contrast,
discrepancies appear in Ni -- a classical itinerant ferromagnet.
This can be attributed to the lack of spin fluctuations in
\emph{GW} diagrams.
However we find out that 
there is no need to include finite-energy spin fluctuations, 
instead
a static correction to the QSGW self-energy is sufficient to correct for the size of the local moment.
While this finding is completely new within such an extensive formalism, it opens up an 
avenue to test the validity of a similar argument for other
transition metals. The LDA or LDA+DMFT should be
problematic, as nonlocality in the self-energy can be important
(see supplemental material).

\subsection{Fe in the Fermi liquid regime}

\begin{figure}
\includegraphics[height=4.0cm,trim=35mm 52mm 20mm 120mm, clip]{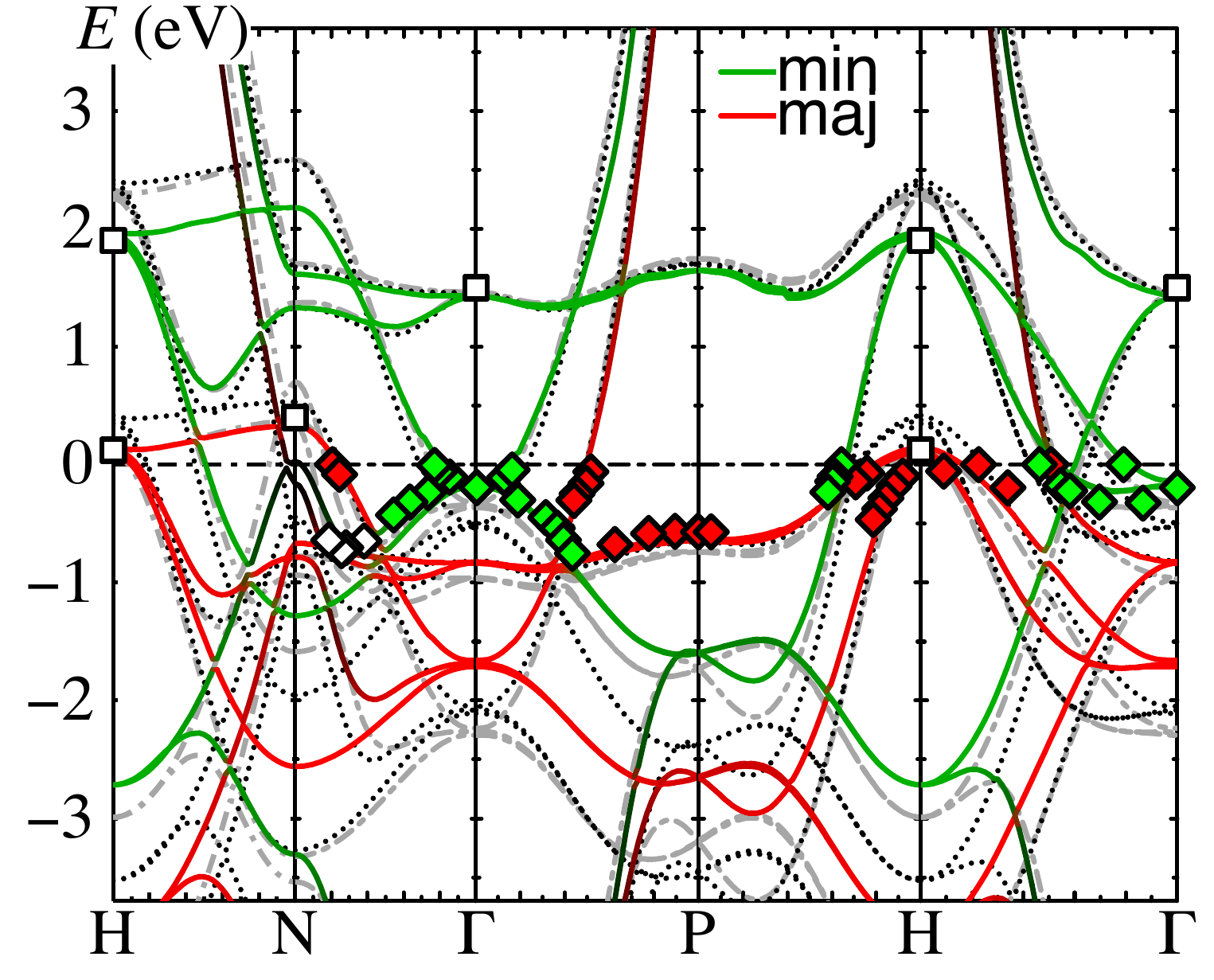} \
\includegraphics[height=4.0cm,scale=0.6]{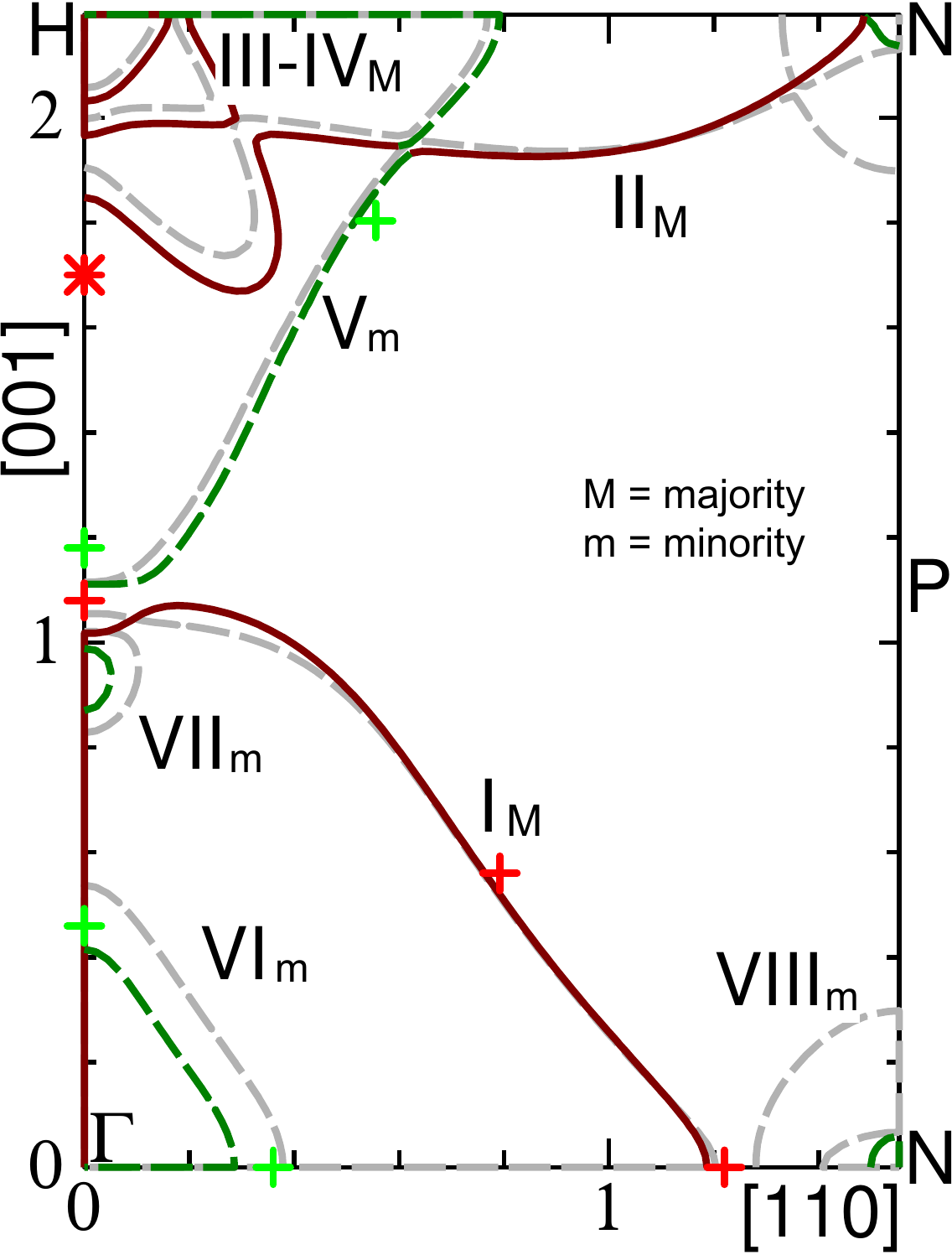}
\caption {\footnotesize \raggedright {(color online) Left: QS\emph{GW} band
         structure of Fe (solid lines), LSDA (grey dashed),
         k-point averaged QS\emph{GW} (black dotted, see text),
         ARPES spectra~\onlinecite{Schafer05} (diamonds) and inverse photoemission
         spectra~\onlinecite{Himpsel91} (squares).
         Right: Fermi surface.  Symbols denote FS crossings reported in
         Ref.~\onlinecite{Schafer05}.  Red and green depict
         majority and minority $d$ character, respectively.}}
\label{fig:QSGWbands}
\end{figure}

\begin{figure}
\includegraphics[height=4.0cm]{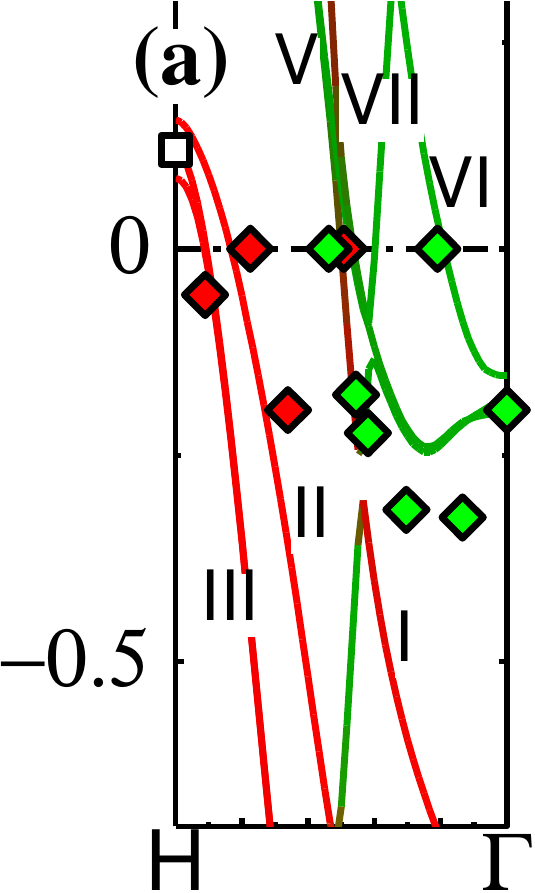} \
\includegraphics[height=4.0cm]{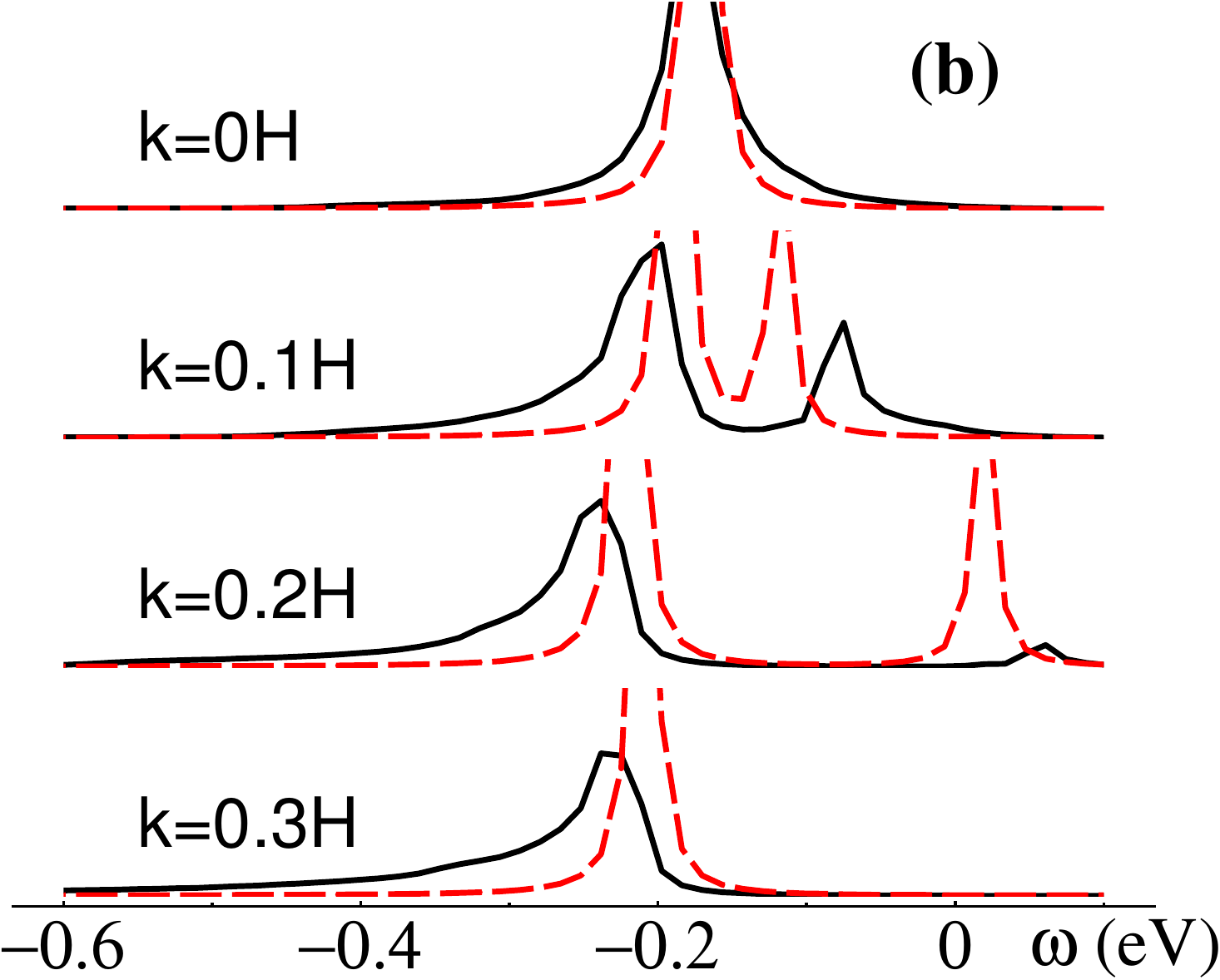} \\ \vskip 6pt
\includegraphics[height=2.5cm]{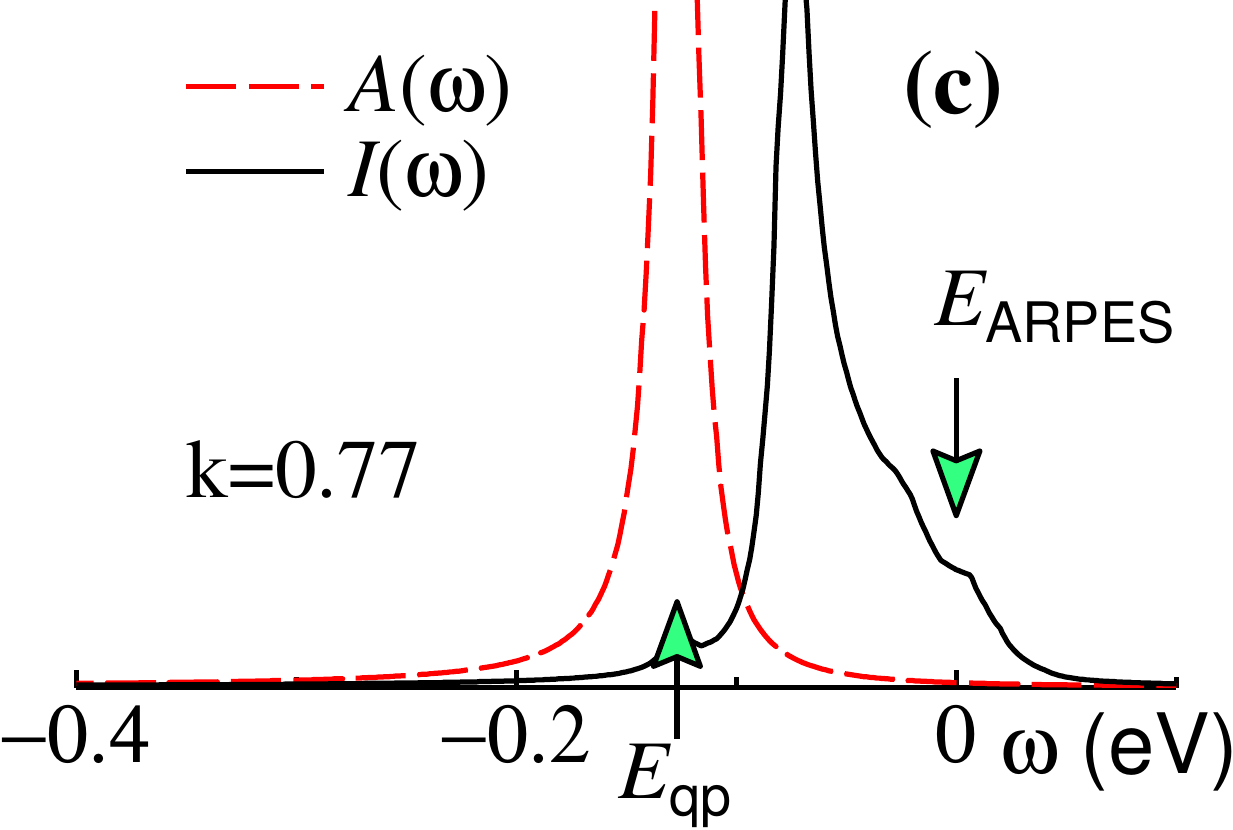} \
\includegraphics[height=2.5cm]{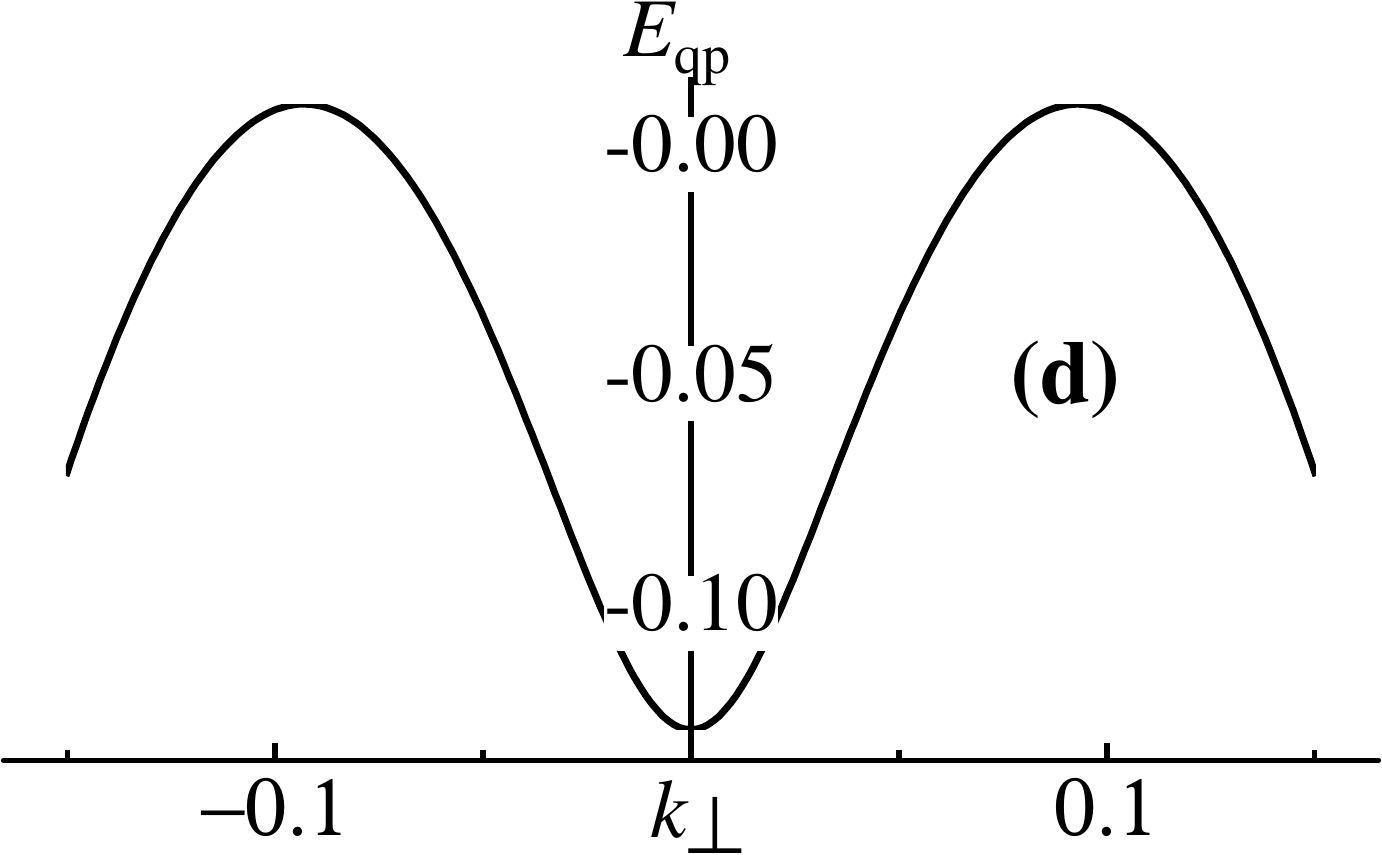} \\ \vskip 6pt

\caption {\footnotesize \raggedright
{ {({\bf a})}: The $\Gamma${-}{\sf H} line of Fig.~\ref{fig:QSGWbands}
  in high resolution.  Labels correspond to traditional assignments of
  Fermi surface pockets \cite{Baraff73,Schafer05}. {({\bf b})}: Dashed line
  is QS\emph{GW} spectral function $A(k,\omega)$ for various points on
  $\Gamma${-}{\sf H} line, with {\sf k}=0 and {\sf k}=1 denoting
  $\Gamma$ and {\sf H}. Solid line is $A(\omega)$ modified according to
  Eq.~(\ref{eq:pes}).
  {({\bf c})}: the analog of {($b$)} at $k{=}0.77\times${\sf H}
  where the {\sf II}$_{\rm M}$ band crosses $E_{F}$. $E_{\rm QP}$ indicates
  the QS\emph{GW} QP level, and $E_{\rm ARPES}$ the experimental
  ARPES peak at 0.77{\sf H}. {({\bf d})}: dispersion in the
  QS\emph{GW} {\sf II}$_{\rm M}$ band on a line
  $\bf{k}_{\perp}$\,+\,[0,0,0.77{\sf H}]
  normal to the film surface.
}}
\label{fig:GamH}
\end{figure}

Fig.~\ref{fig:QSGWbands} compares the calculated QS\emph{GW} band
structure of Fe to peaks in ARPES spectra of Ref.~\cite{Schafer05}, along
with some inverse photoemission data \cite{Himpsel91}.  While agreement appears to be
very good, there are some discrepancies, particularly along the
$\Gamma${-}H line (see also Fig.~\ref{fig:GamH}(a)).  
As
noted earlier, the QS\emph{GW} band structure reflects
the peaks of $A(\bf{k},\omega)$ with no renormalizations
from the $\omega$- or $\bf{k}$- dependence of $\Sigma$.

In the FL regime, ARPES spectra $I(\bf{k},\omega)$ are generally thought
to be a fairly direct measure of $A(\bf{k},\omega)$. But the two are not
identical even in the FL regime, independently of the
precision of the experimental setup.
Assuming a one-step model~\cite{pendry1976theory} for the photoemission
process (initial and final state coupled through Fermi's Golden
rule \cite{matzdorf1998investigation,pendry1976theory})
$I(\bf{k},\omega)$ can be written as
\begin{eqnarray}
\label{eq:pes}
I(\mathbf{k},\omega) \propto \int dk_\perp
|T_{fs}|^{2} |M_{fi}(\mathbf{k}_\bot)|^{2} A_f(\mathbf{k}_\bot) A(\mathbf{k},\omega) ,\\
\hbox{where  }
A_f(\mathbf{k}_\bot) = \frac{{\Delta k_\bot/2\pi}}
{{(\Delta k_\bot/2)^2  + (k_\bot - k_\bot^0)^2 }} \nonumber
\end{eqnarray}
is the spectral function of the final state, broadened by scattering of the
photoelectron as it approaches the surface \cite{strocov2003intrinsic}.
$T_{fs}$ is the final-state surface transmission amplitude and $M_{fi}$ the
photoexcitation matrix element (taken to be constant and
$\mathbf{k}$-independent~\cite{strocov1998absolute}). Thus the final state is
considered to be a damped Bloch wave, taking the form of a Lorentzian
distribution centred in $k^0_\perp$ and
broadened by $\Delta\mathbf{k}_\bot$~\cite{strocov2003intrinsic}, while the initial state is
an undamped Bloch function with an energy broadening
$\Delta E$, obtained through the QS\emph{GW} spectral function.
This approximation is reasonable since in the FL regime
$A(\mathbf{k},\omega)$ is sharply peaked around the QP level.  $\Delta
\mathbf{k}_\bot$ is directly related to the inverse of the electron mean
free path.  For photon energy in the range 100-130\,eV, $\Delta
\mathbf{k}_\bot \approx 0.2~$\AA$^{-1}$
\cite{feibelman1974photoemission,tanuma1994calculations}.

The final-state scattering broadens $I(\omega)$; but it also can shift the
peak $\bar\omega$ in $I(\omega)$.  The most significant discrepancy between
ARPES and QS\emph{GW} is found in the {\sf V}$_{\rm m}$ band,
Fig.~\ref{fig:GamH}($a$) between $k{=}0$ and 0.4$\times${\sf H}.
Fig.~\ref{fig:GamH}($b$) shows $A({\bf{k}},\omega)$ calculated
by QS\emph{GW}, and the corresponding $I({\bf{k}},\omega)$ calculated from Eq.~(\ref{eq:pes}).
Estimating the peak shift change from
$\delta\bar\omega {=} \int d\omega\,\omega I/\int d\omega I - \int
d\omega\,\omega A/\int d\omega A$, we find $\delta\bar\omega {<} 0.01$~eV
at $\Gamma$, increasing to $\delta\bar\omega{\approx}0.06$\,eV for \emph{k}
between 0.1{\sf H} and 0.3{\sf H}.  $\delta\bar\omega$=0.06\,eV tallies
closely with the discrepancy between the {\sf V}$_{\rm m}$ band and the
measured ARPES peak for 0.1{\sf H}${<}k{<}$0.3{\sf H}.  There is also a significant discrepancy in the {\sf
II}$_{\rm M}$ band near $k{=}0.77\times${\sf H}.  Where it crosses $E_{F}$,
the QS\emph{GW} bands deviate from the ARPES peak by nearly 0.15\,eV.
But ARPES simulated by Eq.~(\ref{eq:pes}) is much closer to experiment
(Fig.~\ref{fig:GamH}($c$)).  This is understood from
Fig.~\ref{fig:GamH}($d$), which plots the QS\emph{GW} dispersion along
a line $\Delta\mathbf{k}_{\perp}$ normal to the film surface, passing through [0,0,0.77{\sf H}].
A measurement that includes contributions from this line
biases the ARPES peak in the direction of $E_{F}$ since E$_{\rm qp}$ is
minimum at ${k}_{\perp}{=}0$.  Thus we attribute most of the discrepancy
in the Fermi surface crossing (red star in \ref{fig:QSGWbands}($b$)) to an artifact
of final-state scattering.

To better pin down the errors in QS\emph{GW}, we turn to
de Haas-van Alphen (dHvA) measurements.  Extremal areas of the FS
cross sections can be extracted to high precision
from dHvA and magnetoresistance experiments.
Areas normal to [110] and [111] are
given in Table~\ref{tab:dHvA}, along with areas calculated by
QS\emph{GW}.  Fig.\ref{fig:QSGWbands} shows the QS\emph{GW} Fermi surface, which
closely resembles the one inferred by Lonzarich (version B)
\cite{Lonzarich80}.
There is some ambiguity in
resolving the small VIII$_{\rm m}$ pocket at N because
its tiny area is sensitive to computational details.
Discrepancies in the extremal areas are not very meaningful: it
is more sensible to determine the change $\Delta E_F$ in Fermi
level needed to make the QS\emph{GW} area agree with dHvA
measurements. This amounts to the average error in the
QS\emph{GW} QP levels, assuming that the bands shift rigidly.
This assumption is well verified for all pockets, except for the small VI one owing to strong electron-phonon coupling~\cite{suppmat}.

\begin{table}
\begin{ruledtabular}
\begin{tabular}{lllr@{\hspace*{6pt}}|llr}
FS    &\multicolumn{3}{c}{dHvA [110]} &\multicolumn{3}{c}{dHvA [111]} \\
pocket & QS\emph{GW}  & expt\cite{Baraff73} &  $\Delta E_F$ & QS\emph{GW}  & expt\cite{Baraff73} &  $\Delta E_F$ \\
\hline
I   & 3.355  & 3.3336   &  0.01  & 3.63   & 3.5342 &  0.04\\
II  &        &          &        & 3.694  &        &  \\
III &0.2138  & 0.3190   &  0.05  & 0.1627 & 0.2579 &  0.06   \\
IV  &0.0897  & 0.1175   &  0.04  & 0.0846 & 0.1089 &  0.02   \\
VI  &0.3176  & 0.5559   & -0.13  & 0.2799 & 0.4986 & -0.14 \\
VII & 0.0148 & 0.0405   &  0.04 \\
\hline
\hline
      &\multicolumn{3}{c}{\emph{m}*/\emph{m} [110]} &\multicolumn{3}{c}{\emph{m}*/\emph{m} [111]} \\
          & QS\emph{GW} &     LDA        &  expt\cite{Gold71}
                                         & QS\emph{GW}  & LDA   & expt\cite{Gold71} \\
I         &  2.5        &  2.0   & 2.6    \\
V         &             &        &       &  -1.7  &  -1.2   &  -1.7 \\
VI        &             &        &       & \ 2.0  &\ 1.5    & \ 2.8 \\
\end{tabular}
\end{ruledtabular}
\caption{\raggedright
{\footnotesize
de Haas-van Alphen measurements of extremal areas $A$ on the
[110] and [111] Fermi surfaces, in \AA$^{-2}$.
$\Delta E_F$ is an estimate of the error in the QP level (eV), as described
in the text.  Bottom panel: cyclotron mass,
$m^*/m{=}(\hbar^2/2{\pi}m)\,\partial{A}/\partial{E}$.}}
\label{tab:dHvA}
\end{table}

Some limited cyclotron data for effective masses are also
available~\cite{Gold71}, which are expected to be more reliable
than ARPES data. It is seen that agreement is excellent (Table~\ref{tab:dHvA}, bottom panel)
except for the small VI pocket.
We get a better comparison by accounting for the electron-phonon coupling with
a simple model~\cite{suppmat}.
From the model, $v_F$ is renormalized by a factor $1{+}\lambda{=}1.6$, which reasonably
accounts for discrepancy between the QS\emph{GW} and the cyclotron
mass in pocket VI.
The other pockets are much larger (Fig 1($b$)), making
$v_{F}$ much larger on average, and the renormalization smaller.

Such a perfect agreement with experiments could not be possible 
without the accurate description of non-local components in the QS\emph{GW} self-energy.
To prove this statement we computed the band structure with a local potential
obtained from a k-point average of the QS\emph{GW} self-energy.
The result is reported as a dotted black curve in Figure \ref{fig:QSGWbands},
to be compared with the pale grey lines of LSDA and the solid lines of QS\emph{GW}.
The k-averaged potential reproduces a band structure that is much closer to the LSDA one than to the QSGW results.
This results in the overestimation of the binding energy, e.g. of most states close to Fermi (for instance at $\Gamma$),
or in the range between -2 and -3 eV (see at $\Gamma$, P and H). 

An additional verification that local physics is not relevant in the description of the quasiparticle structure of Fe can be found 
in the Supplemental Materials~\cite{suppmat}. 

\subsection{Ni: an archetypical itinerant magnet}

\begin{figure}[b]
\includegraphics[height=4.0cm]{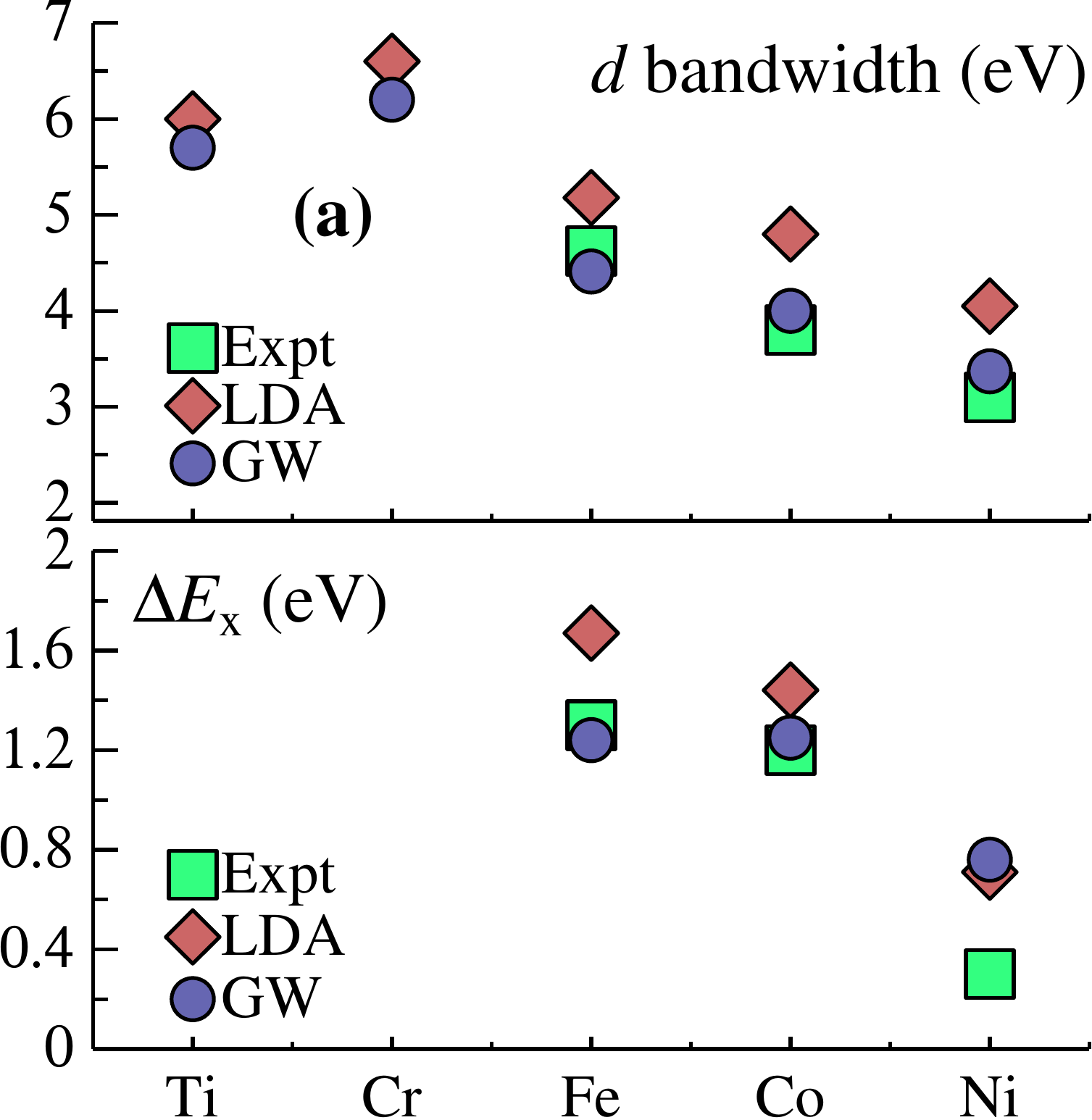} \
\includegraphics[height=4.0cm]{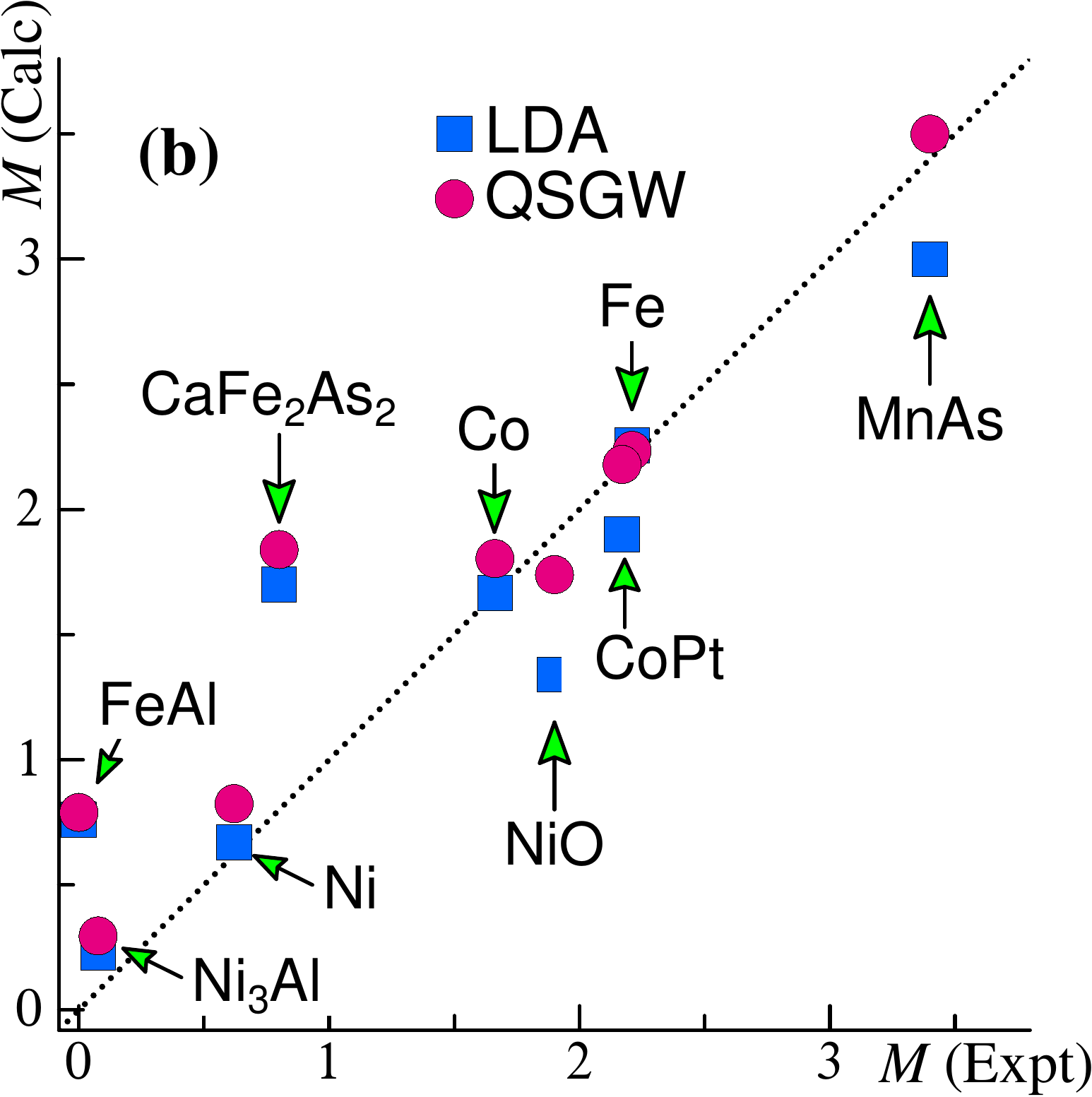}
\caption {\footnotesize \raggedright
{($a$): $d$ bandwidth (top panel) and exchange splitting $\Delta E_x$ (bottom panel) in the 3$d$ elemental metals.
($b$): Magnetic moment of several compounds}}
\label{fig:mmom}
\end{figure}

\begin{figure}
\includegraphics[width=0.49\textwidth]{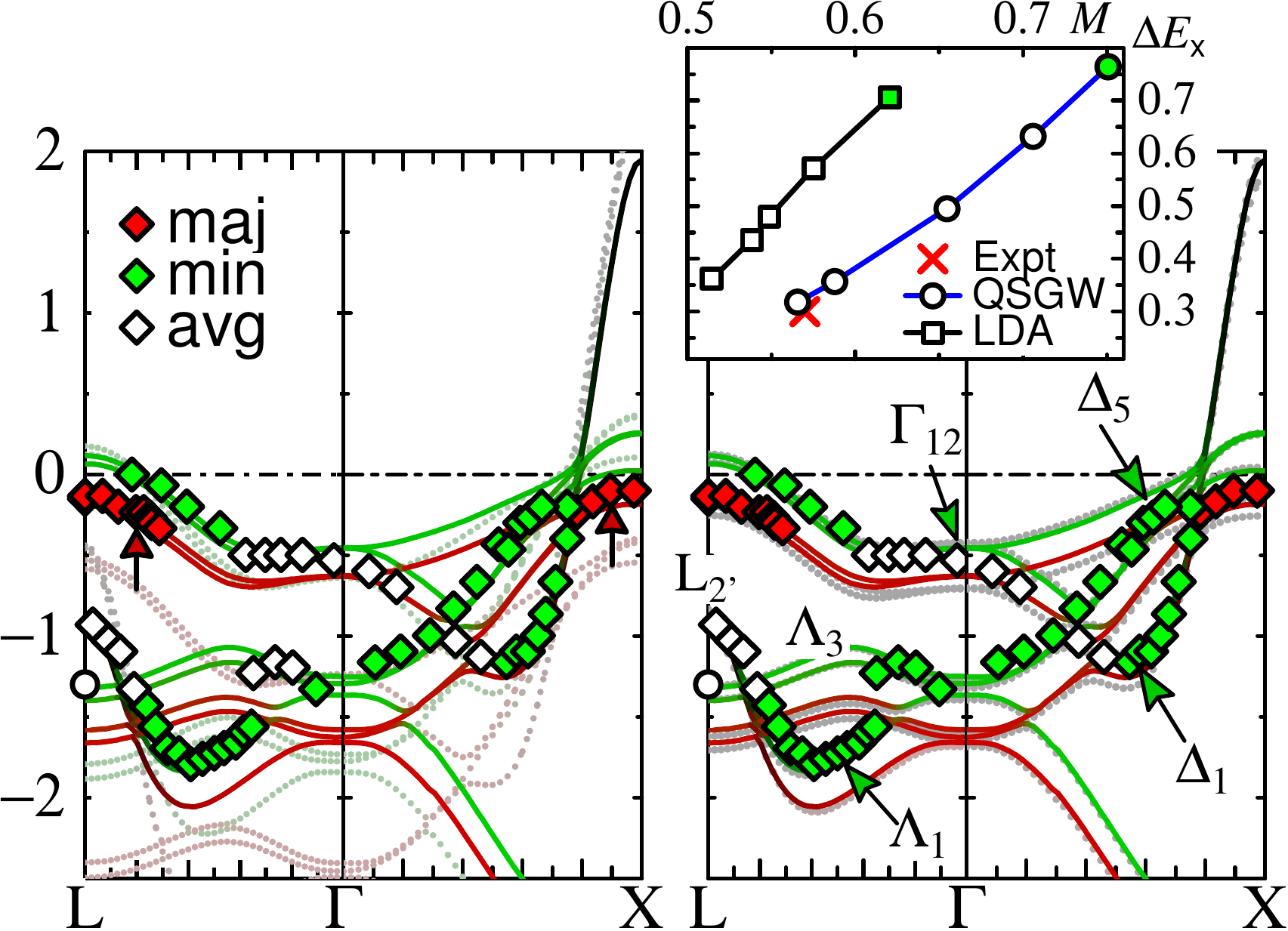}
\caption {\footnotesize \raggedright {(color online) (\emph{left})
 Band structure of Ni in QS\emph{GW} (solid lines) and LDA
    (dotted) and ARPES data \cite{HimpselNi} (the circle at
    $-1.3$\,eV was taken from Ref.~\cite{Eberhardt80}).  Red
    arrows highlight the discrepancy in the exchange splitting
    $\Delta E_x$ at near L and X.  (\emph{right})
    QS\emph{GW}+DMFT bands with method $S$ (solid) and
    QS\emph{GW}+$B^\mathrm{eff}$ (dashed). (\emph{inset}) $\Delta E_x$ at L as a
    function of $M$ obtained by adding an external magnetic field
    to the QS\emph{GW} or LSDA potential (see text).}}
\label{fig:MvsE}
\end{figure}

Less detailed information is available for other elemental transition
metals. We have extracted some experimental bandwidths, and also the
exchange splitting $\Delta E_x$ in the magnetic elements.
Fig.~\ref{fig:mmom}(a) shows that both seem to be very well described
by QS\emph{GW}, except that $\Delta E_x$ deviates strongly from
experiment in Ni.  QS\emph{GW} significantly improves not only on the
LSDA, but also on fully self-consistent \emph{GW}~\cite{Belashchenko06}
because of loss of spectral weight in fully self-consistent $G$
that is avoided in QS\emph{GW}~\cite{Kotani07}.

Fig.~\ref{fig:MvsE}($a$) compares the QS\emph{GW} band structure of Ni
to ARPES data~\cite{HimpselNi}.  Agreement is excellent in the
minority channel, but $\Delta E_x$ is uniformly too
large on the symmetry lines shown.  Also the band near $-1$\,eV
at L (consisting of $s$ character there) is traditionally assumed
to be a continuation of the $d$ band denoted as white and green
diamonds; but the calculations show that at it is a
continuation of Ni $s$ band.  The corresponding LSDA band
(light dotted lines) crosses L at
$E_{F}{-}0.44$\,eV; also the $d$ bands are much wider.


$\Delta E_x$ is about twice too large in both QS\emph{GW} and the
LSDA, and for that reason spin wave frequencies are also too
large~\cite{Karlsson00a}.
Spin fluctuations $\left<M^2\right>$ 
are important
generally in itinerant magnets but they are absent in both LSDA and QS\emph{GW}.
One important property they
have is to reduce the average magnetic moment $\left<M\right>$
and hence to quench $\Delta E_x$ \cite{mazinreview,Shimizu81}.
Fig.~\ref{fig:mmom}($b$) shows this trend quite clearly:
systems such as Fe, Co, and NiO are very well described by
QS\emph{GW}, but $M$ is always overestimated in itinerant magnets
such as FeAl, Ni$_{3}$Al, and Fe based superconductors such
as BaFe$_{2}$As$_{2}$.  Ni is also itinerant to some degree
(unlike Fe, its average moment probably disappears as
$T{\rightarrow}T_{c}$), and its moment should be overestimated.
This is found to be the case for QS\emph{GW}, as
Fig.~\ref{fig:mmom}($b$) shows.

\begin{table}[b]
\centering
\begin{ruledtabular}
\begin{tabular}{lcc}
                              &  $M$ (Bhor) & $\Delta E_x$ @ L (eV)\\  
	\hline
	LSDA                      &    0.62     &  0.71\\
	QS\emph{GW}               &    0.75     &  0.77\\
	QS\emph{GW}+DMFT          &    0.51     &  0.30$\dagger$\\	
	QS\emph{GW}+SLDMFT        &    0.53     &  0.30\\
	QS\emph{GW}+$B^{\rm eff}$ &    0.57     &  0.32\\
	\hline
	Experiment                &    0.57     &  0.31
\end{tabular}
\end{ruledtabular}
\caption {\footnotesize \raggedright {
    Magnetic moment $M$ and exchange splitting $\Delta E_x$ 
    for the different levels of the theory (see text) against experiment.
    $\dagger$Value estimated from Maximum entropy analytic continuation.}}
\label{tab:dmft_results}
\end{table}

Local spin fluctuations are well captured by localized non
perturbative approaches such as DMFT. We can
reasonably expect that the addition of spin-flip diagrams to
QS\emph{GW} would be sufficient
to incorporate these effects.
A $G_0W_0$+DMFT study of ferromagnetic Ni
can be found in \cite{ferdiNi},
but the dependency of $G_0W_0$ on the starting point, 
together with all the advantages mentioned at the beginning, motivated us to devise 
a QS\emph{GW}-based approach.

Here we adopt a novel
implementation merging QS\emph{GW} with DMFT.
We adapted Haule's Continuous Time Quantum Monte Carlo solver
with the projection and embedding schemes described in Ref.~\cite{haule_prb2007},
and which are outlined in the supplemental material~\cite{suppmat}.

The fully consistent QS\emph{GW}+DMFT calculation is composed by alternately repeated
DMFT and QS\emph{GW} loops.
First the QS\emph{GW} Green's function is converged at fixed density, then it is projected on the Ni $d$-orbitals and finally, within the DMFT loop, the local self-energy is obtained.
Updating the total density with the locally corrected Green's function and repeating the procedure leads to complete self-consistency.
This method fully takes into account the dynamics of the local spin-fluctuations included in the DMFT diagrams.
Results are reported in Table~\ref{tab:dmft_results} and they confirm that DMFT adds the correct local diagrams missing in the QS\emph{GW} theory.
Moreover by carefully continuing the resulting Green's function on the real-frequency axis, we find an exchange splitting of $\sim$0.3 eV and a satellite at $\sim$5 eV~\cite{suppmat}.

%

In order to investigate the importance of the dynamics in the local spin-spin channels, we carry out a QS\emph{GW}+DMFT calculation
by retaining only the static limit of the DMFT loop (we call it QS\emph{GW}+``SLDMFT'').
Once the DMFT loop converges, we take the zero frequency limit of the
magnetic part of the DMFT self-energy and add it to the spin-averaged QS\emph{GW} Hamiltonian~\cite{suppmat}. 
As it is clearly shown in Table \ref{tab:dmft_results} and Figure
\ref{fig:MvsE}, this static hamiltonian reproduces very accurately
magnetic moment and details of the band
structure.
This is a strong indication that for Ni the dynamics of local spin
fluctuations is not crucial.  This will be the case if
the quasiparticle picture is a reasonable description of Ni,
even if QS\emph{GW} alone does not contain enough physics to yield an
optimum quasiparticle approximation. 

To verify this further, we model spin fluctuations by carrying out the QS\emph{GW}
self-consistent cycle in the presence of a magnetic field $B^{\rm eff}$,
and tuning $B^\mathrm{eff}$ to reduce $M$ (see inset, Figure \ref{fig:MvsE})
Our key finding is that
when $B^{\rm eff}$ is tuned to make $M$ agree with experiment, $\Delta E_x$
does also, reproducing ARPES spectra to high precision
in the FL regime, as clearly reported in Figure \ref{fig:MvsE} and Table \ref{tab:dmft_results}.
Both the QS\emph{GW} and LSDA overestimate $M$ for
itinerant systems, but the latter also
\emph{underestimates} it in local-moment systems (Fig.~\ref{fig:mmom}($b$)).
In the LSDA treatment of Ni, these effects cancel and render the moment fortuitously good.  When spin
fluctuations are folded in through $B^{\rm eff}$, the LSDA moment becomes too small.
This finding must be interpreted as 
a sign of the superior level of internal consistency in the QS\emph{GW} theory with respect to LSDA.
Without such a degree of consistency spin fluctuations could not be approximated by a static field.

\subsection{Conclusions}

We have performed detailed QS\emph{GW} calculations
of the electronic band structure of several 3$d$ metallic compounds
to assess the reliability of this theory in the Fermi liquid regime 
and the importance of the non-local terms in the self-energy.


\medskip

\emph{\--- \textbf{Fe:} }
Through de Haas-van Alphen and cyclotron measurements we
established that QS\emph{GW} QP levels at $E_{F}$ have an error of
$\sim$\,0.05~eV, and effective masses are well described.
Comparable precision is found below $E_{F}$ by
comparing to ARPES data, provided final state scattering is taken
into account.  The QS\emph{GW} $d$ bandwidth falls in close
agreement with ARPES, and is approximately $0.75$
times that of the LDA (Fig.~\ref{fig:QSGWbands}).

If $\Sigma$ is $k$-averaged to simulate a local self-energy, the
QS\emph{GW} band structure changes significantly and resembles
the LDA. Thus non-locality in the self-energy is important in
transition metals, and its absence explains why LDA+DMFT does not
yield good agreement with ARPES ~\cite{LichtensteinFe09}.

\medskip

\emph{\--- \textbf{Ni:} }
QS\emph{GW} $d$ bandwidths, the
$t_{2g}{-}e_{g}$ splitting, the $s{-}d$ alignment, are all in
excellent agreement with experiment, while $\langle{M}\rangle$
and $\Delta E_x$ are too large.
However through the addition of a uniform static external field 
 QS\emph{GW} can give both in good agreement,
 indicating a high level of consistency in the theory, 
 contrary to LSDA in which is not possible to have both quantities correct at the same time.

To account for spin
fluctuations in an \emph{ab initio} framework, we constructed a
novel QS\emph{GW}+DMFT implementation 
and we utilised it at different degrees of approximations
demonstrating that in itinerant magnets as Ni
(i) the dynamics of fluctuations is irrelevant 
(ii) their effect can be very well
approximated by a static field as long as the non-local correlation part is treated accurately.

\medskip

\begin{acknowledgments}
This work was supported by the Simons Many-Electron Collaboration,
and EPSRC grant EP/M011631/1.
The authors gratefully acknowledge computer resources from the Gauss Centre for Supercomputing e.V. (www.gauss-centre.eu).
We also acknowledge the Partnership for Advanced Computing in Europe (PRACE) for awarding us access to the following resources: Curie FN and TN based in France at the Tr\`{e}s Grand Centre de Calcul (TGCC), and SuperMUC, based in Germany at Leibniz Supercomputing Centre.  
\end{acknowledgments}

%
%
%
%
%
%

\section{Supplemental material}

\subsection{Survey on the QS\emph{GW} theory}
Quasiparticle self-consistency is a construction that determines the
noninteracting Green's function $G_{0}$ that is minimally distant
from the true Green's function $G$.  A measure of distance, or
metric is necessary; a good choice \cite{Kotani07} results in an
effective static potential:
\begin{eqnarray}
\bar{\Sigma}^{\rm xc} = \frac{1}{2}\sum_{ij} |\psi_i\rangle
       \left\{
   {{\rm Re}[\Sigma({\varepsilon_i})]+{\rm Re}[\Sigma({\varepsilon_j})]}_{ij}
 \right\}
       \langle\psi_j|.
\label{eq:veff}
\end{eqnarray}
$\Sigma_{ij}(\omega)$ is the self-energy in the basis of
single-particle eigenstates $|\psi_i\rangle$, which becomes $i G_0 W$
in the \emph{GW} approximation.  Starting from a trial $G_0$, e.g. the
LDA, $\bar{\Sigma}^{\rm xc}$ is determined through \emph{GW}, which determines a
new $G_0$.  The cycle is repeated until self-consistency.

Recently
Ismail-Beigi showed that Eq.~\ref{eq:veff} also minimizes the gradient
of the Klein functional, $|\delta F|^2$, where $F$ is evaluated in the
subspace of all possible static $\bar{\Sigma}^{\rm xc}$ \cite{Ismail14}.

Another key property of Eq.~\ref{eq:veff} is that, at
self-consistency, the poles of $G_0(\mathbf{k},\omega)$ coincide with
the peaks in $G(\mathbf{k},\omega)$ .  Therefore the band structure
generated by $V^{\rm xc}$ coincides with the peaks of the specral
function $A(\mathbf{k},\omega)$.  This is significant, because it
means there is no many-body ``mass renormalization'' of the
noninteracting hamiltonian.  In other words, the attribution of mass
renormalization to correlation effects, a concept widely used in the
literature~\cite{Walter10}, is ill-defined: it depends on an arbitrary
reference, e.g. the LDA.  The absence of mass renormalization is a
very useful property: we can directly associate QS\emph{GW} energy
bands $E(\bf{k})$ with peaks in the spectral function
$A(\bf{k},\omega)$.

\subsection{Electron-phonon renormalization of effective masses}
In comparing the areas of electron pockets in the Fermi surface of Fe, 
we pointed out the small discrepancy between measured and simulated values.
As a measure of the discrepancy we provide the rigid shift of the Fermi energy that would lead the computed area to equate the measured one  (see Table II in main text).
This measure is justified under the assumption that QS\emph{GW} shift rigidly,
which is actually not the case for pocket VI, for which the agreement is poorer.



According to a
Thomas-Fermi model of screening~\cite{Ashcroft76}, the elecron-phonon
interaction renormalizes $v_F$ by a factor $1{+}\lambda$.
Band VI
is roughly spherical, enabling us to evaluate $\lambda$ analytically:
\begin{equation}
\lambda = \frac{e^2}{\hbar v_F}\left[ \frac{1}{2}\ln\frac{k_{TF}^2}{k_{TF}^2+k_{F}^2} +
                                     \frac{k_{F}}{k_{TF}}\arctan\frac{k_{F}}{k_{TF}}\right]
\end{equation}
Estimating $k_F{=}1.71$~\AA$^{-1}$ from the Fe electron
density, this leads to a renormalization factor of $1.6$.

Remembering that $v_F \propto 1/m*$, 
we can compare  
this factor with the ratio $m^*_{\rm QSGW}/m^*_{\rm exp} = 1.4 $, 
which is close to the estimated contribution from the
electron-phonon interaction.

\subsection{Computational details}

\subsubsection{QS\emph{GW}}
For the high resolution needed here, computational
conditions had to be carefully controlled. 

In both QS\emph{GW} calculations of Fe and Ni,
a $k$ mesh of
12$\times$12$\times$12 divisions was found to be sufficient for calculating
$\Sigma$.  The one-body part was evaluated on a 24$\times$24$\times$24
mesh.

Fe 3\emph{p} and 4\emph{d} states were included through local
orbitals: omitting these and treating 3$p$ as core levels \cite{Kotani07}
can shift QP levels by as much as 0.1 eV in the FL regime.  Other
parameters \cite{Kotani07}, such as broadening the pole in $G$ in
constructing $\Sigma{=}iGW$, the basis of eigenfunction products, and the
energy cutoff for the off-diagonal parts of $\Sigma$, were also carefully
monitored.  When set to tight tolerances QP levels near $E_{F}$ were stable
to a resolution of 0.05 eV.  QP levels are calculated including spin-orbit
coupling (SOC), though it is omitted in the calculation of $\Sigma$.
The effect of SOC on $\Sigma$ was found to make small changes to $\bar{\Sigma}^{\rm xc}$.

Similar parameters were used in the QS\emph{GW} calculation for Ni.

\subsubsection{Our QS\emph{GW}+DMFT implementation}
Concerning the QS\emph{GW}+DMFT calculation on Ni,  
we projected the lattice problem on the Ni $d$ orbitals 
following the prescription of Haule~\cite{kriproj}.
We compute higher level diagrams locally using the hybridization
expansion version of the numerically exact continuous
time QMC method~\cite{werner_prl2006,haule_prb2007}.

In order to single out 
the correlated subspace, 
a procedure of projection/embedding 
which was originally introduced in~\cite{kriproj} in the LAPW basis of the Wien2k package, 
is developed in the Full-Potential Linear Muffin-Tin Orbitals (FP-LMTO) basis~\cite{FPLMTO}. 
This projector maps the full space Green's function  $G_{ij{\bf k}}$
(with band and $k$-point label $\{ij{\bf k}\}$) 
to the local Green's function $G_{LL'}^\mathrm{loc}$ defined only on the correlated subspace.
The compact index $L\coloneqq \{\tau,R,\sigma,\ell,m \}$, collects information on
the atom type $\tau$, site $R$, spin coordinate $\sigma$, and angular momentum components $\ell$ and $m$.
The projection operation can be cast in the following form:
\[
G_{LL'}^\mathrm{loc} = \sum_{{\bf k},ij} U^L_{i\bf k} \,G_{ij{\bf k}}\, {U^{L^\prime}_{j \bf k}}^\dagger
\quad\text{with}\quad 
U^L_{i\bf k} \propto \sum_u \mathcal{A}^L_{i\bf k} \Phi^u_{R \ell} \;, 
\]
where the coefficients $ \mathcal{A}^L_{i\bf k}$ account for localization inside the sphere,
while $ \Phi^u_{R \ell}$ gives an estimate of correlations relative to the specific orbital component $\ell$.
More specifically $ \mathcal{A}^L_{i\bf k}$ are linear combinations 
of spherical harmonics $Y_{\ell m}$ and the QS\emph{GW} quasiparticle eigenfunctions 
in the FP-LMTO basis. 
The terms $\Phi^u_{R \ell}$ 
are radial integral of the kind 
$\braket{\varphi^u_{R\ell}}{\varphi_{R\ell}}$ 
where the index $u$ in $\varphi^u_{R\ell}$ indicates the possibility of selecting 
the radial solution $\varphi_{R\ell}$ of the Schr\"{o}dinger equation inside the MTO, its energy derivative $\dot{\varphi}_{R\ell}$, and its local orbitals contributions $\varphi^z_{R \ell}$~\cite{Kotani07}. 
By means of these definitions we ensure that the localized orbitals 
are centred on the correlated atom 
corresponding to the muffin-tin site $R$. 

The transformation matrices $U$ have been orthonormalised in such a way that
$\sum_{i{\bf k}}{U^{L}_{i \bf k}}{U^{L'}_{i \bf k}}^\dagger = \delta_{LL'}$.

The local Green's function is defined on a grid of Matsubara frequencies $i\omega_n=i\pi(2n+1)/\beta$ and it is employed to calculate the hybridization function of the system, which feeds the CTQMC impurity solver.
The result of the impurity solver is the local impurity self-energy $\Sigma^{\rm loc}_{LL'}(i\omega_n)$ also defined on the Matsubara axis. In order to update the full Green's function $G_{ij{\bf k}}$ with this local self-energy, so to iterate the DMFT loop to self-consistency, an embedding procedure is needed.
Because of the specific properties of the 
transformation $U$,  
the embedding procedure 
$\Sigma^{\rm loc}_{ij{\bf k}}(i \omega_n) = \sum_{LL'} U^{L\dagger}_{i{\bf k}} \Sigma_{LL'}^{\rm loc}(i\omega_n) U^{L'}_{j{\bf k}} $
can be operated by means of the same matrices,
even though this is not a general requirement of the theory~\cite{kriproj}.

The charge double-counting contribution has been included by means of
the standard formula 
\[ E_{dc} = U(n-1/2) - J(n/2-1/2) \]
where $n$ is the nominal occupancy of the $3d$ shell.

\subsubsection{Analytical continuation through Maximum entropy}

\begin{figure}
	\centering
	\includegraphics[width=0.40\textwidth, trim=20mm 20mm 20mm 10mm, clip]{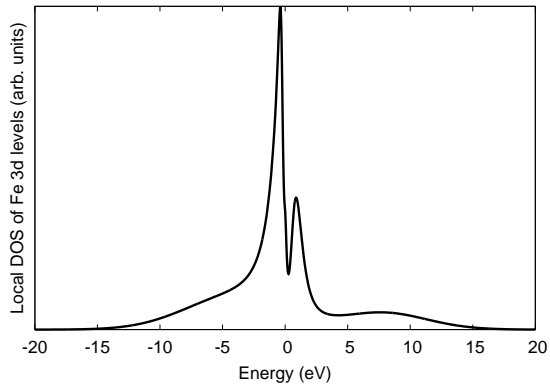}
	\caption{Total DOS of Fe 3$d$ states obtained from Maximum entropy method.}
	\label{fig:Fe_spf}
\end{figure}

The output of the DMFT loop is the impurity self-energy $\Sigma^{\rm loc}_{LL'}(i\omega)$ and the corresponding impurity Green's function $\mathcal{G}_{LL'}(i\omega)$. 
At self-consistency they correspond to the local self-energy and local Green's function of the correlated subset
($d$-electrons of the metal in this case).

Though, since the CTQMC solver works in the Matsubara's frequency space
extrapolations to zero-energy or some kind of analytical continuation has to be employed to extract physically meaningful quantities.

We evaluated the scattering rate from the intercept at $\omega=0$ of a forth-order polynomial obtained from a fit of $\text{Im}\Sigma^{\rm loc}_{LL'}(i\omega)$ in the vicinity of $\omega=0$.
The coefficient $C_1$ of the linear term is then related to the quasiparticle weight $\mathcal{Z}=1/(1+C_1)$.
The intercept being $\Gamma/\mathcal{Z}$ where $\Gamma$ is the scattering rate, we have been able to evaluate $\Gamma$ between $10^{-3}$ and $10^{-4}$ for both Fe and Ni, corresponding to a coherent charge-scattering regime.

The spectral function of the correlated subset can be obtained by continuing on the real-frequency axis the quantity $A(i\omega)=-\text{Im}G_{LL'}(i\omega)/\pi$.
In order to do that, we utilised a statistical technique based on the Maximum Entropy Method~\cite{gubernatis-jarrel_prb1991}.
As model function we used the magnetic QS\emph{GW} spectral function of the $d$ orbitals.

The resulting spectral function $A(\omega)$ contains all the effects of the dynamical local response including features beyond the quasiparticle peaks, as satellites at higher energy. This is indeed the case for Fe, reported in Figures~\ref{fig:Fe_spf}.

\subsubsection{Applications to Ni}

In our application to Ni,
the projectors used for the Ni $3d$ are constructed from 5 bands
below $E_{F}$ and 3 bands above $E_{F}$, which correspond to a
window of ${\sim}{\pm}10$~eV.  By choosing a wide energy
window, $U$ becomes nearly static~\cite{Choi15}. The
corresponding on-site Coulomb parameters were chosen to be
$U$=10~eV and $J$=0.9~eV, as calculated by
constrained RPA~\cite{Choi15}.
The Matsubara frequency grid is defined over 2000 points with an inverse 
temperature $\beta=50$~eV$^{-1}$.

In the case of Ni we applied the method at two different degrees of approximation.
\begin{description}

\item[Standard procedure] We first perform a full QS\emph{GW} loop. 
Then we extract the spin-averaged Green's function from which the local hybridization function $\Delta(i\omega_n)$ is obtained.
The DMFT solver uses $\Delta(i\omega_n)$, $U$ and $J$ to produce the local self-energy $\Sigma^{\rm loc}_{LL'}(i\omega_n)$.
By keeping the QS\emph{GW} part of the Green's function fixed, the calculation of the self-energy is repeated until convergence of the magnetic moment (DMFT loop).
Then a new density $\rho(\mathbf{r})$ is recomputed summing over all Matsubara's frenquencies according to
\begin{equation}
	\rho(\mathbf{r}) =   \sum_{\mathbf{k},l} \left\{\frac{1}{\beta}\sum_{n=-\infty}^{+\infty}  
				 \frac{\psi^{R}_{l \mathbf{k}}(\mathbf{r},i\omega_n)\psi^{L*}_{l \mathbf{k}}(\mathbf{r},i\omega_n)}{i\omega_n + \mu -\varepsilon_{l\mathbf{k}}(i\omega_n)} \right\}\,, \label{eq:first_rho}
\end{equation}
where $\psi^{X}_{l \mathbf{k}}(\mathbf{r},i\omega_n)$ are the right ($X=R$) and left ($X=L$) eigenfunctions of the DMFT Hamiltonian $H_{\bf k}(i\omega_n)$ with corresponding eigenvalues $ \varepsilon_{l\mathbf{k}}(i\omega_n)$. 
A new QS\emph{GW} loop is then converged by keeping the density fixed, that produces a new QS\emph{GW} Green's function.
The output is used to initialise a new DMFT loop and so on until QS\emph{GW} self-energy and DMFT updated density are self-consistent.

The number of iterations in each DMFT loop varies between 10 and 20, 
and the number of random moves per iteration are roughly $10^{10}$.
This method is basically the equivalent of state-of-the-art DMFT implementations in the framework of DFT+DMFT.
We believe it is superior 
to other $G_0W_0$+DMFT implementations because of the possibility to close the full self-consistent loop and because of the higher quality of the low-level theory chosen.

\item[QS\emph{GW}+SLDMFT] From the converged DMFT self-energy, we first extrapolate the static limit of $\Sigma^{\rm loc}_{LL'}(i\omega_n)$,
then we embed it into the lattice problem 
and we keep only its symmetrized real part
\begin{equation}
\bar{\Sigma}^{\rm loc}_{ij{\bf k}}=\text{Re}\left[ \Sigma^{\rm loc}_{ij{\bf k}}(0) + \Sigma^{\rm loc}_{ji{\bf k}}(0) \right]/2 .
\end{equation}
This is done for the spin-up and spin-down channels separately. 
We finally retain only the spin-flip component $\Sigma_2=\bar{\Sigma}^{\rm loc}-(\bar{\Sigma}^{\rm loc}_\uparrow + \bar{\Sigma}^{\rm loc}_\downarrow)/2$ and we add it 
to the charge component $\Sigma_1=(\bar{\Sigma}^{\rm xc}_\uparrow + \bar{\Sigma}^{\rm xc}_\downarrow)/2 $ computed at QS\emph{GW} level.
This procedure allowed us to prevent counting twice the magnetic contributions to the self-energy.
\end{description}

\subsubsection{Applications to Fe}
To further confirm that the local momentum is not relevant in the description of Fe, 
we applied the fully dynamical QS\emph{GW}+DMFT approach also to Fe. 
The parameters $U=5$~eV and $J=0.8$~eV used are the same as in\cite{wang_natcomm2013}, 
where they have been derived from GW within the same projection scheme.
Four valence and four conduction bands have been used to project onto the correlated system.
The resulting magnetic moment $M$=2.20 Bohr, as the QS\emph{GW} prediction, confirming that 
 local corrections are absolutely negligible in Fe.

 
We extract the scattering rate $\Gamma$ from the intercept and the linear term of a forth-order polynomial fitting Im$\Sigma$($\omega$) in the vicinity of $\omega=0$ as explained above.
We find that Fe has extremely coherent charge scattering.
We also present the analytically continued ImG(i$\omega$)/$\pi$ for Fe reported in Figure~\ref{fig:Fe_spf}.





\end{document}